# WiFi Assisted Multi-WiGig AP Coordination for Future Multi-Gbps WLANs

[1,2]Ehab Mahmoud Mohamed, [1]Hideyuki Kusano, [1]Kei Sakaguchi, and [1]Seiichi Sampei
[1]Graduate School of Engineering, Osaka University, Osaka, Japan.
[2]Electrical Engineering Dept., Aswan University, Aswan, Egypt.
Email: ehab@wireless.comm.eng.osaka-u.ac.jp, {sakaguchi, sampei}@comm.eng.osaka-u.ac.jp

*Abstract*—Wireless Gigabit (WiGig) access points (APs) using 60 GHz unlicensed frequency band are considered as key enablers for future Gbps wireless local area networks (WLANs). Exhaustive search analog beamforming (BF) is mainly used with WiGig transmissions to overcome channel propagation loss and accomplish high rate data transmissions. Due to its short range transmission with high susceptibility to path blocking, a multiple number of WiGig APs should be installed to fully cover a typical target environment. Therefore, coordination among the installed APs is highly needed for enabling WiGig concurrent transmissions while overcoming packet collisions and reducing interference, which highly increases the total throughput of WiGig WLANs. In this paper, we propose a comprehensive architecture for coordinated WiGig WLANs. The proposed WiGig WLAN is based on a tight coordination between the 5 GHz (WiFi) and the 60 GHz (WiGig) unlicensed frequency bands. By which, the wide coverage WiFi band is used to do the signaling required for organizing WiGig concurrent data transmissions using control/user (C/U) plane splitting. To reduce interference to existing WiGig data links while doing BF, a novel location based BF mechanism is also proposed based on WiFi fingerprinting. The proposed coordinated WiGig WLAN highly outperforms conventional un-coordinated one in terms of total throughput, average packet delay and packet dropping rate.

## I. INTRODUCTION

Wireless gigabit (WiGig), which is standardized by IEEE 802.11ad [1] and IEEE 802.15.3c [2], indicates the multi-Gbps wireless transmissions using the 60 GHz millimeter wave unlicensed frequency band. IEEE 802.11ad defines the use of multi-band (2.4, 5 and 60 GHz) access points (APs) for backward compatibility with the previous IEEE 802.11 a, b, g, and n standards and for fast session transfer (FST) among them. FST is introduced for enabling multi-band WiGig stations (STAs) to transfer the ongoing data transmission from one band to another band based on the radio signal strength of the different bands. IEEE 802.11ad also introduces single medium access control (MAC) address (virtual MAC address) for both WiFi (2.4/5 GHz) and WiGig (60 GHz) to facilitate the FST operation. Due to its high operating frequency, WiGig transmissions suffer from high propagation loss and shadowing effect. Thus, antenna beamforming (BF) is mainly used to combat these channel impairments and accomplish data transmissions [3]. MAC based BF protocols are mainly used as appropriate BF mechanisms for WiGig transmissions. In which, exhaustive search, using switched antenna array with a structured codebook, is used to estimate the best TX/RX beam directions for an AP-STA link [4] [5].

As a consequence of WiGig short range transmissions and its high susceptibility to path blocking, a multiple number of WiGig APs should be installed to fully cover a typical target environment using WiGig concurrent transmissions [6]. Also, it provides the potential to enhance the robustness against path blocking through handover between the installed APs. WiGig concurrent transmissions can be effectively done through utilizing spatial multiplexing inherent in WiGig directional transmissions [6]. However, installing a numerous number of autonomously operated WiGig APs, with no coordination among them, adds several challenges come from the exhaustive search BF protocol. In the widely used random access scenario using CSMA/CA, the carrier sense (CS) function of WiGig AP might not indicate the medium busy due to the predominant nature of directional transmissions and receptions. Accordingly, nearby APs may start the exhaustive search BF simultaneously, or an AP may perform the exhaustive search BF while its nearby APs are involved in directional data transmissions. Both cases cause a lot of packet collisions and interference if the APs are autonomously operated. Consequently, the total system performance in terms of total system throughput, packet dropping rate and average packet delay will be highly degraded. As a result, increasing the number of WiGig APs may result in degrading the total system performance if there is no coordination among the deployed APs. The current standardized architectures for WiGig networks only consider point-to-point and point-to-multi-point use cases [1] [2]. Neither system architecture nor MAC protocol are considered for the case of multi-point-to-multi-point WiGig networks, which is the key enabler for future high capacity multi-Gbps WLANs. Accordingly, a new system architecture and a new MAC protocol are highly needed to efficiently coordinate WiGig concurrent transmissions in random access scenario as an extension to the current WiGig standards.

In this paper, a comprehensive network architecture is proposed for coordinated WiGig WLANs. Moreover, a novel dual-band (5, 60 GHz) MAC protocol using CSMA/CA is proposed to achieve the WiGig coordination operation. To facilitate the coordination function among the dual-band APs, a tight coordination between WiFi and WiGig bands is proposed, which can be considered as a more general WiFi/WiGig coordination compared to the FST proposed by IEEE 802.11ad. As well, several technologies are proposed in accordance to this WiFi/WiGig tight coordination.

The first proposed technology is an introduction of control/user (C/U) plane splitting [7] over the proposed dual-band MAC protocol. In the proposed C/U plane splitting, control frames to be shared among the APs are transmitted via the wide coverage WiFi band, while the high speed data frames are concurrently transmitted by the WiGig band from the multiple APs. Through using this technology, all control signaling including CS are done in the 5 GHz band using its omni-directional wide coverage signals, which highly overcomes the problems of WiGig CS come from its directional transmissions. Control frames transmitted by WiFi signaling include frames needed to coordinate the BF training among the APs in the WiGig band. Thus, only one AP can perform the BF training at a time. Hence, packet collisions due to simultaneous BF can be effectively eliminated. Also, not to harm an existing WiGig data link, the link information such as the used beam identification (ID), modulation coding scheme (MCS) index and received power, is broadcasted using WiFi signaling. Consequently, other APs can effectively exclude beam IDs (bad beams) that may interfere with this existing data link from their BF training beams.

The second proposed technology is to use WiFi fingerprinting to estimate the best and bad beams IDs of the WiGig links [8]. The idea behind this technology is that we can roughly identify the location of a user equipment (UE) by using WiFi channel information measured by multiple APs that is called fingerprint in this paper. Since the best AP to be associated and the best WiGig beam ID to be selected are location dependent, the exhaustive search BF training is not needed anymore if we have a database (DB) to make link between WiFi fingerprints and WiGig best beam IDs. For that purpose, offline statistical learning is introduced in this paper. Based on this statistical learning DB, in real-time BF, by just comparing current UE WiFi fingerprint readings with the pre-stored UE WiFi fingerprints, a best associated AP can be selected for the UE, and a group of WiGig best sector IDs (beams) can be estimated for the selected AP to effectively communicate with the UE at its current position. Among these estimated best beams, the beam IDs overlapping with the existing WiGig data links are recognized as bad beams and eliminated from the BF refinement process. These bad beams can be easily estimated using the pre-constructed statistical learning DB.

Using computer simulations, the proposed coordinated WiGig WLAN highly outperforms the conventional un-coordinated one in terms of total system throughput, average packet delay and packet dropping rate.

The rest of this paper is organized as follows; Section II provides the proposed coordinated WiGig WLAN including the proposed dual band MAC protocol. The proposed WiGig BF based on WiFi fingerprinting is presented in Sect. III. The performance of the proposed WiGig WLAN is analyzed in Sect. IV via computer simulations. Section V concludes this paper.

## II. WiFi Assisted Multi-WiGig AP Coordination

Since the WiFi alliance has integrated the WiGig alliance, the future WLAN chipsets must support multi-band such as 2.4, 5, and 60 GHz. In accordance, IEEE 802.11ad introduces the FST, which works effectively only in the case of single AP where handover is occurred only between the homothetic coverage of WiFi and WiGig. In the case of multiple APs, handover between APs, as well as avoiding interference between the selected antenna beams of multiple APs should be considered; of which functions cannot be afforded by the current FST mechanism [1]. In this section, we will introduce a more comprehensive coordination between WiFi and WiGig bands to realize future high capacity multi-Gbps WLANs.

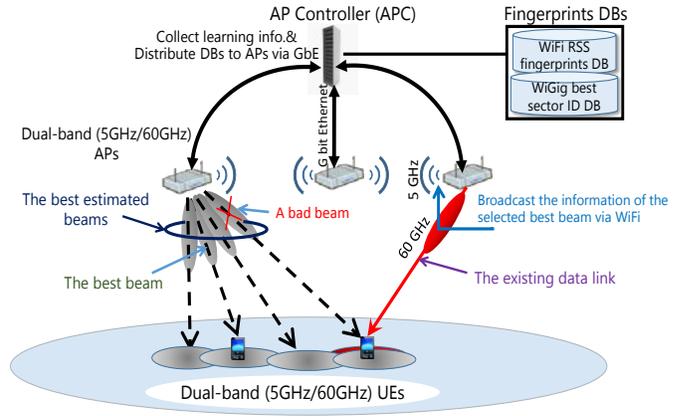

Fig. 1. Sub-cloud WLAN system architecture.

### A. Proposed Sub-cloud Network Architecture for Coordinated WiGig WLAN

Figure 1 shows the system architecture of the proposed sub-cloud WiGig WLAN using WiFi/WiGig tight coordination. In Fig. 1, multiple dual-band (5 and 60 GHz) APs are connected to an AP controller (APC) via gigabit Ethernet links. This sub-cloud WLAN will be installed in a target environment to cover it using multiple dual-band APs. In this architecture, the APC works as a central coordinator of the sub-cloud to assist radio resource management among the APs. Likewise, the APC acts as a gateway to connect the WLAN to the Internet. The 5 GHz WiFi band in Fig.1 is mainly used to broadcast the signaling information required for enabling 60 GHz WiGig concurrent transmissions with low packet collisions and interference.

To prevent packet collisions result from simultaneous BF training of multiple APs, the AP which intends to do BF starts the random access process using the WiFi band to prevent any other AP from doing BF until it finishes. Thus, only one AP can do the BF training at a time. Also, not to interfere with existing WiGig data links while doing BF, the AP excludes the beam directions (bad beams) that may interfere with existing links from the training beams, as it is shown in Fig. 1. In this paper, to effectively estimate the bad beam IDs, statistical learning is used.

In the proposed statistical learning, the APC collects WiFi and WiGig best sector ID fingerprints at arbitrary learning points (LPs) in the target environment, and it stores them as fingerprints DBs, see Fig.1. In this paper, we use WiFi received signal strength (RSS) as a simple fingerprint of the WiFi signal; other WiFi fingerprinting methodologies can be easily used without any modifications to the general approach [9]. Fingerprints collection is done in an offline phase, which is not repeated unless the transmit power or locations of the APs are changed, or the internal structure of the environment is changed. Then, the APC performs grouping and clustering on the collected WiFi fingerprints to find out the best exemplars for each WiGig best sector ID.

In the online phase, after collecting and comparing current UE WiFi RSS readings with the pre-stored WiFi RSS exemplars, the APC associates the UE to an appropriate AP from the un-used ones based on the offline records. A group of best sector IDs (best beams) is also estimated for the selected AP-UE link for beam refinement, as shown in Fig. 1. Therefore, the exhaustive search BF is not needed any more using this statistical learning approach. Instead, only BF refinement using the high data rate beam refinement protocol (BRP) frames is used.

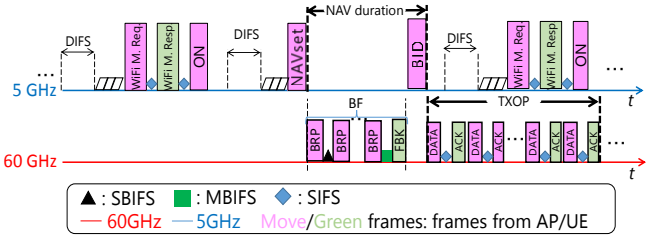

Fig.2. Dual-band MAC protocol for coordinated WiGig WLAN.

By only using beam refinement and removing the sector level sweep (SLS) phase used in the conventional exhaustive search BF, the average packet delay is highly reduced by reducing the BF setup time. Also, it contributes in increasing the total throughput of the proposed WLAN by removing the overhead comes from the SLS phase. Among the estimated best beams, the beam directions that may interfere with the existing links are excluded and recognized as bad beams, as shown in Fig.1. These bad beams are easily estimated using the pre-constructed fingerprint DBs as it will be discussed in more details through the subsequent subsections. Therefore, efficient coordinated BF among multiple APs can be achieved by WiFi/WiGig tight coordination in the proposed sub-cloud WLAN.

The APC can provide other coordination functions like association/re-association between WiFi and WiGig bands of all APs based on the data rate provided by each AP on each band and the user traffic load, which is left as our future work.

### B. Proposed Dual-band MAC Protocol

Figure 2 shows the proposed dual-band MAC protocol, which is used to coordinate WiGig concurrent transmissions in random access scenarios using WiFi signaling. In this protocol, if a data frame is generated for a specific UE, the random access process starts using WiFi interface instead of WiGig. In which, one of un-used APs senses the 5 GHz band using CS routine. If the medium is free, it starts the backoff counter. If the counter reaches zero, it starts to send a WiFi measurement request (WiFi M. Req.) frame to intended UE. Then, UE broadcasts a WiFi measurement response (WiFi M. Resp.) frame for measuring its current WiFi fingerprint. Based on the current WiFi fingerprint and offline fingerprint DBs, the APC associates UE to an appropriate un-used AP. Then, it sends a switch ON frame to the UE from the un-used AP using its WiFi interface to turn ON its WiGig interface if it is in a sleep mode. The APC also estimates a group of best beams for the selected AP-UE link and groups of bad beams to make interference to this link from other APs. Then, the APC sends this best and bad beam information to the coordinated WiGig APs. After this preparation process, the selected AP starts the random access process again in the 5 GHz band. If the backoff counter reaches zero, a NAVset frame is sent from the AP using its WiFi interface to prevent any other AP from doing BF refinement in 60 GHz band until it finishes. The NAVset frame contains the estimated time that the AP will take until it finishes BF refinement. The BF refinement selects the best beam from the group of best beams by using high speed BRP frames. At the end of the BF refinement process, UE sends a feedback (FBK) frame to the AP using the 60 GHz band to inform the ID of the highest link quality beam (the best beam) and its received power. Consequently, the AP broadcasts a best beam identification (BID) frame, which contains the estimated best beam ID, the used MCS index and the received power using WiFi signaling. By broadcasting the BID frame, other APs will consider the selected link information when they accurately estimate their bad beams before conducting BF refinement. After broadcasting the BID, the AP starts to send data frames to the UE using the coordinated best beam in 60 GHz band.

Due to the orthogonality between WiFi and WiGig bands, the WiFi fingerprint measurements can be done during WiGig data transmissions as shown in Fig. 2, which highly reduces the latency in the proposed protocol.

### III. BEST BEAMS ESTIMATION AND BAD BEAMS ELIMINATION USING STATISTICAL LEARNING

To efficiently coordinate the BF process and reduce packet collisions and interference to existing WiGig data links while doing BF, the concept of best and bad beams are introduced. The best beams are estimated based on the current UE WiFi fingerprint (location) by using statistically learned fingerprint DBs. This best beams estimation speeds up the BF training by removing the SLS phase in the exhaustive search BF. Instead, only BF refinement is needed. From these estimated best beams, beam directions (bad beams) that may degrade the data rate of existing WiGig data links are eliminated from the BF refinement. The candidates of bad beams can be estimated by evaluating the overlaps of best beams of different APs in the best beam DB. In the online phase, more accurate bad beams are selected by using the BID frames sent from existing links.

### A. Offline Statistical Learning

The first step in the offline statistical learning phase is to construct the 5 GHz and 60 GHz radio maps for target environments. Constructing the radio maps can be effectively done by collecting the average WiFi RSS readings (fingerprints) and WiGig APs best sectors IDs at arbitrary LPs in the target environment. Therefore, three databases are constructed as radio maps, i.e. the WiFi fingerprint DB $\Psi$, the best sector ID DB $\Phi$, and the offline received power DB $\mathbf{P_{OFF}}$, which are defined as:

$$\Psi = \begin{pmatrix} \psi_{11} & \cdots & \psi_{L1} \\ \vdots & \ddots & \vdots \\ \psi_{1N} & \cdots & \psi_{LN} \end{pmatrix}, \Phi = \begin{pmatrix} \phi_{11} & \cdots & \phi_{L1} \\ \vdots & \ddots & \vdots \\ \phi_{1N} & \cdots & \phi_{LN} \end{pmatrix},$$

$$\mathbf{P_{OFF}} = \begin{pmatrix} p_{11}^{\phi_{11}} & \cdots & p_{L1}^{\phi_{L1}} \\ \vdots & \ddots & \vdots \\ p_{1N}^{\phi_{1N}} & \cdots & p_{LN}^{\phi_{LN}} \end{pmatrix}, \quad (1)$$

where $\psi_{ln}$ is the WiFi fingerprint at AP $n$ from a dual band UE located at LP $l$. $L$ is the total number of LPs, and $N$ is the total number of APs. $\phi_{ln}$ is the WiGig best sector ID at LP $l$, which corresponds to the antenna sector ID of AP $n$ which maximizes the received power by a UE located at LP $l$. $\phi_{ln}$ can be calculated as:

$$\phi_{ln} = d_n^* = arg\max_{d_n}(P_{ln}(d_n)), \quad 1 \leq d_n \leq D_n, \quad (2)$$

where $d_n$ indicates sector ID of AP $n$, $D_n$ is the total number of sector IDs, and $\phi_{ln} = d_n^*$ is the best sector ID at LP $l$ from AP $n$ that maximizes the received power $P_{ln}(d_n)$. A *null* sector ID in the $\Phi$ matrix, i.e., $\phi_{ln}$ = *null*, means that AP $n$ cannot cover LP $l$. The *null* values are used by the APC for association/re-association decisions. $p_{ln}^{\phi_{ln}}$ is the power received at LP $l$ from AP $n$ using best sector ID $\phi_{ln}$. When $\phi_{ln}$ is equal to *null*, $p_{ln}^{\phi_{ln}}$ becomes to 0.

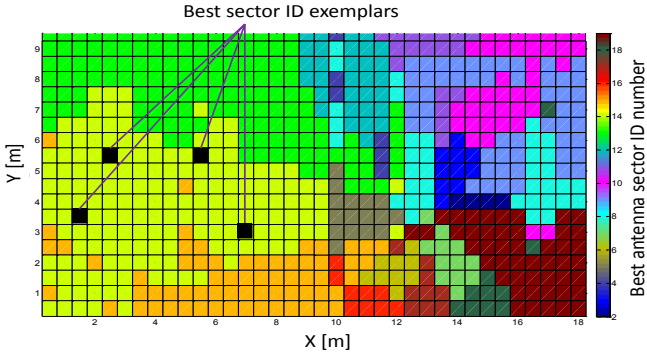

Fig. 3. An example of a typical WiGig AP best sector ID radio map.

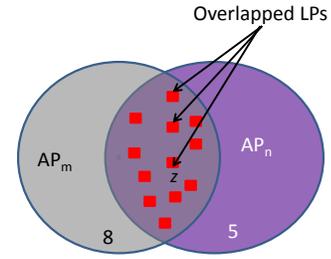

Fig. 4. An example of overlapped LPs that can be covered by AP *n* best sector ID 5 and AP *m* best sector ID 8.

Figure 3 shows an example of a WiGig AP best sector ID radio map using uniformly distributed LPs in a room area of 180 m². In this example, the WiGig AP is located at x = 14 m, y = 3.5 m and z = 3 m. The color bar indicates sector ID. Each square in Fig. 3 represents a different LP, which is covered by a certain best sector ID, and it has a certain WiFi fingerprint.

To effectively localize a best sector ID $d_n^*$ of AP *n* and to reduce the computational complexity of WiFi fingerprint matching in the online phase, grouping and clustering are applied on the WiFi fingerprints. Namely, the WiFi fingerprints corresponding to the same best sector ID $d_n^*$ are grouped together then clustered. Because the WiGig radio map given in Fig. 3 is highly irregular, affinity propagation clustering algorithm [10] is used as an appropriate clustering algorithm to form the clusters. At the end of the clustering operation, we obtain a set of WiFi fingerprint exemplars $\mathfrak{J}_j^{d_n^*}$, $j \in \{1,2,\ldots,C_{d_n^*}\}$ for each best sector ID $d_n^*$, where $C_{d_n^*}$ is the total number of WiFi fingerprint exemplars (clusters) for $d_n^*$. In Fig. 3, four WiFi exemplars are calculated to effectively localize/represent the best sector ID 15 (yellow color), and Fig. 3 shows the LPs corresponding to these exemplars.

### B. Online Best Beam Selection and Bad Beams Elimination

The actual WiGig concurrent transmissions take place in this real-time phase. During this phase, the online WiFi fingerprint vector $\boldsymbol{\psi}_r$ from the target UE, at an arbitrary position *r*, is measured by the APs and collected by the APC. $\boldsymbol{\psi}_r$ is defined as:

$$\boldsymbol{\psi}_r = [\psi_{r1} \psi_{r2} \ldots \psi_{rN}]^T. \quad (3)$$

Using $\boldsymbol{\psi}_r$, $\boldsymbol{\psi}$, $\boldsymbol{\Phi}$ and $\mathbf{P}_{\mathbf{OFF}}$, the APC can associate the UE to an appropriate un-used AP *n*. After that, the APC estimates a group of best beams for AP *n* to communicate with the UE at its current position *r*. This can be done by calculating the smallest Euclidian distance between $\boldsymbol{\psi}_r$ and the WiFi fingerprint exemplars $\mathfrak{J}_j^{d_n^*}$ of each best sector ID $d_n^*$. Therefore, a vector of smallest Euclidian distances is obtained with a length up to the total number of best sector IDs. Then, the APC sorts the obtained vector of smallest Euclidian distances in an ascending order, and it selects a group of best sector IDs (best beams) $d_n^*(1:X)$ up to *X*, where *X* is the total number of estimated best beams, as follows:

$$d_n^*(1:X) = \underset{d_n^*}{\text{sort}} \left( \arg \min_{1 \leq j \leq C_{d_n^*}} \left\| \boldsymbol{\psi}_r - \mathfrak{J}_j^{d_n^*} \right\|^2 \right) \bigg|_{1:X}. \quad (4)$$

After estimating the best beams $d_n^*(1:X)$, the APC pre-estimates bad beam candidates from other APs *m* ($m \neq n$) that may collide/interfere with these estimated best beams.

In evaluating bad beams candidates, we used the following criterion:

$$d_{bm}^{d_n^*(x)}(i) = \forall_{d_m^*, m \neq n} \left( MCS_{d_n^*(x)}^{d_m^*} < MCS_{d_n^*(x)} \right), \quad 1 \leq x \leq X,$$
$$1 \leq m \neq n \leq N, \ 1 \leq i \leq Cand_{bm}^{d_n^*(x)} \quad (5)$$

where $MCS_{d_n^*(x)}$ is the ideal MCS value if AP *n* uses best beam $d_n^*(x)$ without interference. $MCS_{d_n^*(x)}^{d_m^*}$ is the realized MCS value when AP *n* uses best beam $d_n^*(x)$ while AP *m* uses best beam $d_m^*$ at the same time. $d_{bm}^{d_n^*(x)}(i)$ is the ID of bad beam candidate of AP *m* that degrades the MCS of $d_n^*(x)$, and $Cand_{bm}^{d_n^*(x)}$ is the total number of bad beam candidates against $d_n^*(x)$ at AP *m*. The APC calculates (5) based on $SNR_{d_n^*(x)}$ and $SINR_{d_n^*(x)}^{d_m^*}$, where $SNR_{d_n^*(x)}$ is the received signal to noise ratio if AP *n* uses best beam $d_n^*(x)$, $SINR_{d_n^*(x)}^{d_m^*}$ is the signal to interference pulse noise ratio if $d_m^*$ interferes with $d_n^*(x)$. The APC calculates $SNR_{d_n^*(x)}$ and $SINR_{d_n^*(x)}^{d_m^*}$ for every overrllaped LP *z* covered by AP *n* and AP *m* using $d_n^*(x)$ and $d_m^*$ in the $\boldsymbol{\Phi}$ matrix as shown in Fig. 4. The calculation is done based on the $\mathbf{P}_{\mathbf{OFF}}$ as:

$$SNR_{d_n^*(x)}(z) = \frac{p_{nz}^{d_n^*(x)}}{\sigma^2}, \quad (6)$$

$$SINR_{d_n^*(x)}^{d_m^*}(z) = \frac{P_{zn}^{d_n^*(x)}}{P_{mz}^{d_m^*} + \sigma^2}, \quad 1 \leq z \leq Z, \quad (7)$$

where $p_{nz}^{d_n^*(x)}$ and $P_{mz}^{d_m^*}$ are the offline power received at overlapped LP *z* from AP *n* using best beam $d_n^*(x)$ and from AP *m* using best beam $d_m^*$ respectively, $\sigma^2$ is the noise power, and *Z* is the total number of overlapped LPs. If at least one of the overlapped LPs satisfies (5), the ID number of $d_m^*$ is registered as a bad beam candidate from AP *m* to $d_n^*(x)$, i.e., $d_{bm}^{d_n^*(x)}(i)$. After estimating $d_n^*(1:X)$ and $d_{bm}^{d_n^*(x)}$, the APC sends $d_n^*(1:X)$ to AP *n*, and $d_n^*(1:X)$ and $d_{bm}^{d_n^*(x)}$ to AP *m*. After NAV set using WiFi, AP *n* conducts BF refinement to find out the best beam $d_n^*(x^*)$ among $d_n^*(1:X)$, and then it broadcasts its ID along with the actual received power and the actually used MCS index using the BID frame via WiFi signaling. By knowing the actually used values of $d_n^*(x^*)$, $p_{nz}^{d_n^*(x^*)}$ and $MCS_{d_n^*(x^*)}$, other APs can find out more accurate bad beams $d_{bm}^{d_n^*(x^*)}(i)$ using above criterion.

Later, if AP *m* is selected for associating another UE, it will eliminate the estimated bad beams $d_{bm}^{d_n^*(x^*)}$ from its estimated best beams $d_m^*(1:X)$ before starting BF refinement to not interfere with the existing AP *n* link. Thus, a collision free and interference less BF training can be achieved.

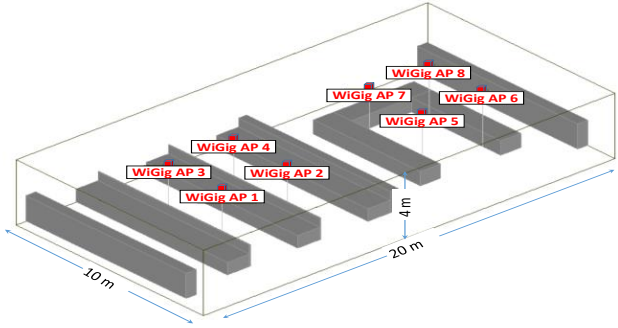

Fig. 5. Raytracing simulation area.

TABLE I. SIMULATION PARAMETERS

| Parameter | Value |
|---|---|
| Transmission mode | SC/FDE |
| Number of APs/UEs | 8/24 |
| Tx power of 5 GHz/60 GHz | 20 dBm/10 dBm |
| Beamwidth in azimuth and elevation directions | 30°, 30° |
| Num. of antenna sectors / beam gain | 36 / 25 dBi |
| Num. of estimated best beams | 6 |
| Num. of LPs | 90 |
| Traffic model | Poisson distribution |
| Offered load / Packet size | 1Gbps / 1500 octet |
| Max. num. of re-transmission | 10 |
| Beacon interval time | 1sec |

## IV. SIMULATION ANALYSIS

In this section, the efficiency of the proposed coordinated WiGig WLAN compared to the conventional un-coordinated one is verified via computer simulations.

### A. Simulation Area and Simulation Parameters

Figure 5 shows a raytracing simulation area in an indoor office environment. We simulate the detailed MAC specifications given in IEEE 802.11ad and IEEE 802.11 standards. Critical simulation parameters are given in Table I. The steering antenna model defined in IEEE 802.11ad [1] is used as the transmit antenna directivity for WiGig AP, in which, the 3D beam gain in dB is given as:

$$G(\varphi,\theta)[dB] = G_0[dB] - \min[-(G_H(\varphi) + G_V(\theta)), A_m], \quad (8)$$

$$A_m[dB] = 12 + G_0[dB], \quad (9)$$

$$G_0[dB] = 20\log_{10}\left(\frac{1.6162}{\sin\left(\frac{\theta_{-3dB}}{2}\right)}\right), \quad (10)$$

where $\varphi, \theta$ are the azimuth and elevation angles. $G_H(\varphi)$, $G_V(\theta)$ are the beam gains in horizontal and vertical directions, which are defined as:

$$G_H(\varphi) = -\min\left[12\left(\frac{\varphi - \varphi_{beam}}{\varphi_{-3dB}}\right)^2, A_m\right], \quad (11)$$

$$G_V(\theta) = -\min\left[12\left(\frac{\theta - \theta_{tilt}}{\theta_{-3dB}}\right)^2, A_m\right], \quad (12)$$

where $\varphi_{beam}$ and $\theta_{tilt}$ are the angles of beam center, $\varphi_{-3dB}$ and $\theta_{-3dB}$ are the half power beamwidths. To fully cover the area, 3D beamforming is used, where each beam has different values of $\varphi_{beam}$ and $\theta_{tilt}$. Using (8), the total channel response from WiGig AP in the 60 GHz band becomes:

$$g(\tau) = \int_0^{2\pi} \int_0^{\pi} \sqrt{G(\varphi,\theta)}\, h(\varphi,\theta,\tau)\sin\theta d\theta d\varphi, \quad (13)$$

where $h(\varphi,\theta,\tau)$ is the channel response without BF gain.

### B. Performance Metrics

In the conducted simulations, we concern in measuring the followings:
- **Total system throughput [Gbps],** which is defined as a sum of throughputs for all successfully delivered packets.
- **Average packet delay,** which is defined as the average time period from the instant when a packet occurs to the instant when the UE completes reception of it.
- **Packet dropping rate [%],** which is defined as: *ND*x100/(*NS* +*ND*), where *NS* denotes the number of successfully received packets, and *ND* denotes the number of discarded packets due to exceeding the re-transmission limit, see Table I.

### C. Simulation Results

Figures 6, 7 and 8 show the average total throughput, the average packet delay and the average packet dropping rate, respectively. In these figures, we compare the performance of the proposed coordinated WiGig WLAN with the conventional un-coordinated one against the number of used APs. The configuration of used APs are summarized in Table II.

TABLE II. SIMULATED AP CONFIGURATION

| Number of APs | AP Configuration |
|---|---|
| 1 | AP 1 |
| 2 | APs 1 and 8 |
| 4 | APs 1, 2, 7 and 8 |
| 6 | APs 1, 2, 3, 4, 5 and 7 |
| 8 | APs 1,2,3,4,5,6,7 and 8 |

In the case of 1 AP, the average packet dropping rate reaches zero because there is no source of packet collisions in both WLANs. However, using the proposed coordinated WiGig WLAN, the average packet delay is decreased compared to the un-coordinated one as shown in Fig. 7. This is because, in the proposed coordinated WLAN, the exhaustive search BF mechanism using the low data rate SLS phase is completely eliminated. Instead, thanks to the proposed statistical learning approach, only a fast BF refinement using the high data rate BRP frames is used. This contributes in reducing the BF setup time and packet overhead compared to the un-coordinated WLAN, which results in not only decreasing the average packet delay but also increasing the average total throughput as shown in Fig. 6.

As we increase the number of used APs, which we try to disperse their locations as much as possible for fair comparisons as in Table II, the average total throughput of the conventional un-coordinated WLAN is not highly increased. Instead, the average total throughput is even decreased when all 8 APs are operated. This comes from the high packet collisions resulting from the autonomously operated exhaustive search BF and from the high interference between the constructed WiGig concurrent data links. These packet collisions also increase the packet dropping rate and the average packet delay. Using 8 APs, the average packet dropping rate is more than 30 % and the average packet delay is more than 350 msec as shown in Figs 7 and 8.

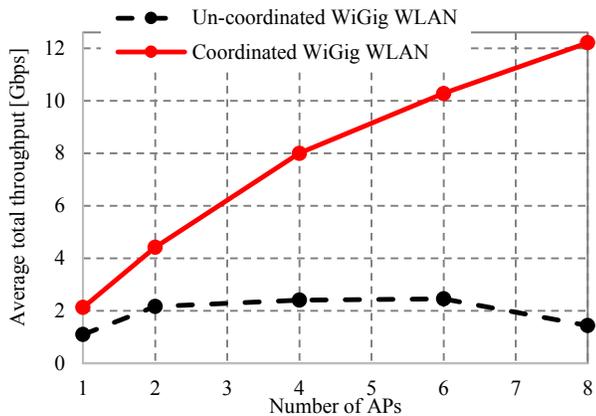

Fig. 6. Average total throughput [Gbps].

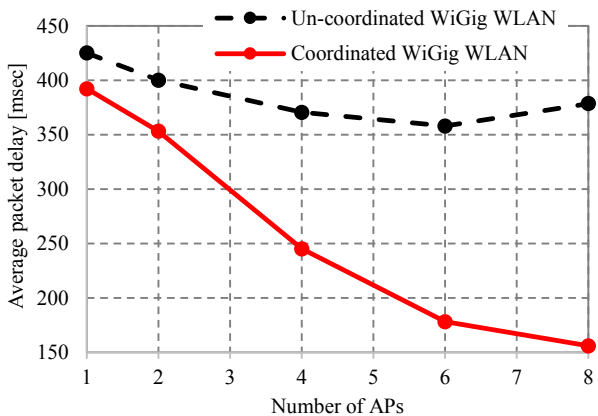

Fig. 7. Average packet delay [msec].

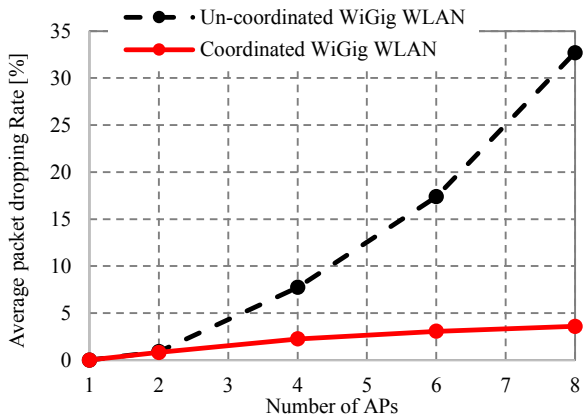

Fig. 8. Average packet dropping rate [%].

On the other hand, using the proposed coordinated WiGig WLAN, as we increase the number of used APs, the average total throughput is highly increased even when all 8 APs are simultaneously operated. As shown in Fig. 6, the average total throughput is almost linearly increased as the number of used APs is increased. This means that only a few number of packets are dropped with a low interference between the established WiGig concurrent links of multiple APs. Due to the inaccuracy in the bad beams estimation, which is mainly based on the pre-stored offline records, some interference still exists between the established WiGig concurrent beams. This especially happens when 6 and 8 APs are used, where the number of APs is high and their coverages are highly overlapped. In consequence, the average total throughput is not linearly increased using 6 and 8 APs as shown in Fig. 6.

V. CONCLUSION

In this paper, we proposed an efficient WiFi/WiGig tight coordination for future multi-Gbps WiGig WLANs. The proposed WiFi/WiGig coordination enables WiGig concurrent transmissions in random access scenarios while reducing packet collisions and interference. The C/U plan splitting is proposed to establish WiGig concurrent data transmissions using the wide coverage WiFi signaling. A low interference BF refinement is also introduced based on the proposed concept of best beam selection and bad beam elimination. The proposed BF refinement is mainly based on the fact that the best associating AP and the best communicating beam are location dependent. In the proposed algorithm, the best beams are selected and the bad beams are eliminated based on the WiFi fingerprint corresponding to that location by using statistically learned database. Moreover, in the proposed dual band MAC protocol, WiFi control frames are used not only for fingerprint measurements, but also to share the selected beam information among the coordinated WiGig APs. We proved that the proposed coordinated WiGig WLAN highly outperforms the un-coordinated one in terms of average total throughput, average packet delay, and average packet dropping rate.

ACKNOWLEDGMENT

This work was partly supported by "The research and development project for expansion of radio spectrum resources" of The Ministry of Internal Affairs and Communications, Japan.